\documentclass[fleqn,twoside]{article}
\usepackage[headings]{espcrc2}

\readRCS
$Id: espcrc2.tex,v 1.2 2004/02/24 11:22:11 spepping Exp $
\usepackage{graphicx}
\usepackage[figuresright]{rotating}

\newcommand{\AmS}{{\protect\the\textfont2
  A\kern-.1667em\lower.5ex\hbox{M}\kern-.125emS}}

\newcommand{\ipb}{\ensuremath{\mathrm{pb^{-1}}}}

\newcommand{\TeV}{\ensuremath{\matqq    hrm{Te\kern -0.1em V}}}
\newcommand{\TeVc}{\ensuremath{\mathrm{Te\kern -0.1em V\!/}c}}
\newcommand{\TeVcc}{\ensuremath{\mathrm{Te\kern -0.1em V\!/}c^2}}
\newcommand{\GeV}{\ensuremath{\mathrm{Ge\kern -0.1em V}}}
\newcommand{\GeVc}{\ensuremath{\mathrm{Ge\kern -0.1em V\!/}c}}
\newcommand{\GeVcc}{\ensuremath{\mathrm{Ge\kern -0.1em V\!/}c^2}}
\newcommand{\MeV}{\ensuremath{\mathrm{Me\kern -0.1em V}}}
\newcommand{\MeVc}{\ensuremath{\mathrm{Me\kern -0.1em V\!/}c}}
\newcommand{\MeVcc}{\ensuremath{\mathrm{Me\kern -0.1em V\!/}c^2}}

\newcommand{\um}{\ensuremath{\mathrm{\mu m}}}

\newcommand{\psinv}{\ensuremath{\mathrm{ps^{-1}}}}
\newcommand{\cdfii}{CDF\,II~}
\newcommand{\ppbar}{p\overline{p}}
\newcommand{\at}{\symbol{64}}

\newcommand{\myto}{\kern -0.3em\to\kern -0.2em}

\newcommand{\Dms}{\ensuremath{\Delta m_s}}
\newcommand{\Dmd}{\ensuremath{\Delta m_d}}

\newcommand{\kstar}{\ensuremath{K^{\ast 0} }}

\newcommand{\Bs}{\ensuremath{B_s^0}}
\newcommand{\Bd}{\ensuremath{B_d^0}}
\newcommand{\BsDspi}{\ensuremath{\bar{B_s^0} \myto D_s^+ \pi^-}}
\newcommand{\BslDs}{\ensuremath{\bar{B_s^0} \myto D_s^+ l^- X}}

\newcommand{\DsPhipi}{\ensuremath{D_s^+ \myto \phi \pi^+}}
\newcommand{\DsKstarK}{\ensuremath{D_s^+ \myto \bar{K}^{\ast 0} K^+}}
\newcommand{\DsTrepi}{\ensuremath{D_s^+ \myto \pi^+ \pi^- \pi^+}}
\newcommand{\Phipi}{\ensuremath{\phi \pi^+}}
\newcommand{\KstarK}{\ensuremath{\bar{K}^{\ast 0} K^+}}
\newcommand{\Trepi}{\ensuremath{\pi^+ \pi^- \pi^+}}

\title{B meson mixing at \cdfii}

\author{M. Rescigno (for the CDF Collaboration)\address[Roma]{Istituto Nazionale di Fisica Nucleare, Sezione di Roma, P.le Aldo Moro 2, 00185 Roma, Italy}}
       
\begin{document}

\begin{abstract}
We present the first limit on \Bs\ mixing frequency obtained using 
360 \ipb\ of Tevatron Run II data with the \cdfii detector. 
We derive $\Dms > 7.9\ \psinv @\ 95\%$ C.L. with a sensitivity 
of 8.4\psinv. 
\vspace{1pc}
\end{abstract}

\maketitle

\section{Introduction}

In the Standard Model (SM) flavor mixing of neutral mesons 
is induced by second-order charged weak interactions
through box diagrams. For \Bd\  
the mixing frequency, \Dmd, is $\propto f^2_{B_d}B_{B_d}|V_{td}|^2$
and thus provides a constraint on the usual unitarity triangle
whereas for \Bs\ $\Dms\propto f^2_{B_s}B_{B_s}|V_{ts}|^2$ 
and is much higher since $|V_{ts}| >> |V_{td}|$.
Theoretical uncertainty on the hadronic parameters make the constraint
from the precise \Dmd\ measurement weak. 
On the other end a measurement of the mixing frequency ratio 
$\Dms/\Dmd$ would fix the ratio of  
hadronic parameters thus allowing a much tighter constraint on 
the unitarity triangle. The expected range for \Dms\ from the 
analysis of all available B and Kaon physics data within the SM
is getting narrower and narrower 
($\Dms \in [15.6, 23.1]\ \at\ 95\% {\rm C.L.}$~\cite{UTFIT}).
Observing \Bs\ mixing outside the SM favored range
would be a signal for new physics. Experiments at LEP and SLD 
as well CDF in Run I have only provided a lower limit 
$\Dms > 14.5 \at 95\%$ C.L.~\cite{pdg2004}.
The high frequency expected for \Bs\ flavor oscillations 
poses formidable experimental challenges related to the need for
extremely accurate proper time resolution in order 
to resolve the oscillation pattern. This is clearly reflected
in the approximate expression for the significance of the
mixing amplitude measurement~\cite{Moser}
\begin{equation}
\label{eq:significance}
{\cal S}= \frac{A}{\sigma_A} \simeq
 \frac{S}{\sqrt{S+B}}\sqrt{\frac{\varepsilon D^2}{2}} 
                     e^{-\frac{1}{2}(\sigma_{c\tau}\Dms)^2}
\end{equation}
where $S$ is the signal sample size, $B$ is the background, 
$\varepsilon$ is the efficiency for flavor tagging and $D$
is the tagging dilution defined as $\frac{N_R-N_W}{N_R+N_W}=1-2w$,
where the mistag rate $w=N_W/(N_R+N_W)$ is the fraction 
of B candidates with an incorrectly assigned flavor. 
The effect of finite resolution 
on proper time ($\sigma_{c\tau}$) reduces the statistical 
significance of the measurement, especially at high values of
\Dms. A key point of the CDF upgrade for Run II has been the 
capability of triggering on hadronic decays of $B$ mesons
allowing the collection of fully reconstructed \Bs\ decays. 
Despite the limited statistics 
available in these modes the excellent proper time 
resolution significantly helps the current sensitivity 
of CDF and will offer the unique chance of a determination 
of \Dms\ from CDF data alone. 
We describe here the first search for \Bs\ mixing using 
360 \ipb\ of Run II data.

\section{\cdfii Detector and Trigger}

The upgraded CDF detector for Run II~\cite{CDFII} 
has been designed to enhance the sensitivity
to \Bs\ mixing, in particular with the L00~\cite{L00} layer 
of silicon detector very close to the beam pipe,
the TOF system~\cite{TOF} and the Silicon Vertex Tracker 
(SVT)~\cite{SVT}. This latter device, reconstructing on-line tracks
in the silicon tracker, enabled triggering on 
$B$-decay vertices, for the first time at a hadron collider.
The current CDF $b-$physics related trigger menu is based on 
the following selections:
\begin{itemize}
\item Di-muons with muon transverse momenta $P_T > 1.5\ \GeVc$ 
to select $b\rightarrow J/\psi X$ decays.
\item Single muon or electron with $P_T>4$ associated with
a displaced trigger track with impact parameter greater than 120 \um\
(lepton+SVT). 
\item A Two Track Trigger (TTT) requiring two displaced SVT
tracks with impact parameters greater than 120 \um, consistent with 
originating at a vertex 200 \um\ away from the beam axis.
\end{itemize}
These different triggers provide large samples of, respectively, 
$B_{d,u}\myto J/\psi K^{(*)}$, 
$B_{d,u}\myto l D^{(*)}X$ and  $B_{d,u}\myto D^{(*)}\pi$
used to calibrate the vertex resolution function, 
the flavor-tagging dilution and to extract the impact parameter 
trigger bias. 
These measurements are an essential part of the $B_s$ mixing study
making the latter really the ``ultimate'' $b$-physics analysis 
at CDF.

\section{Signal samples}

\begin{figure}[t]
\vspace{9pt}
\includegraphics[width=0.87\linewidth]{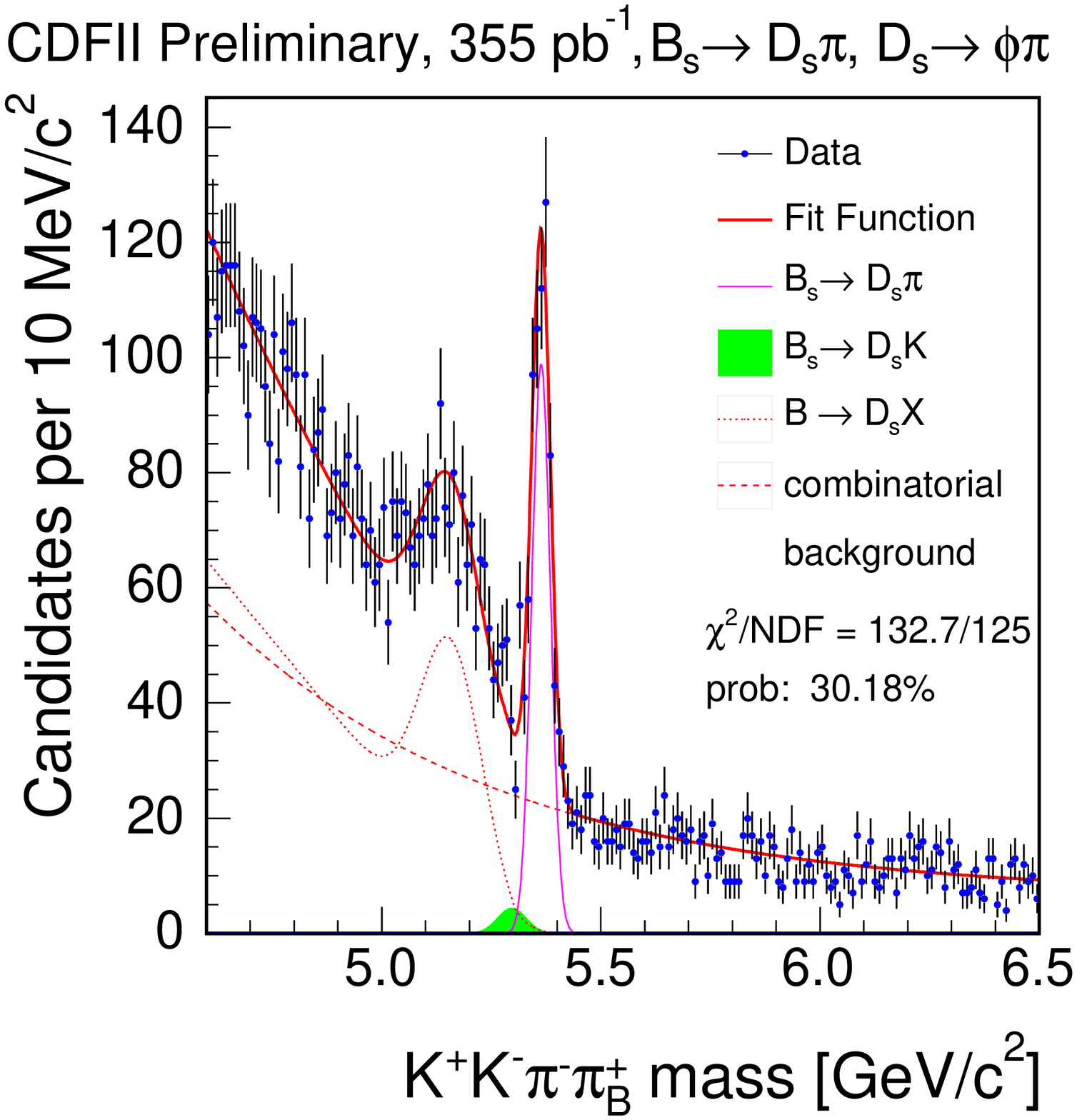}
\vspace{-27pt}
\caption{\BsDspi,\DsPhipi signal.}
\label{fig:Bshadronic}
\end{figure}

\begin{figure}[t]
\vspace{9pt}
\includegraphics[width=.98\linewidth]{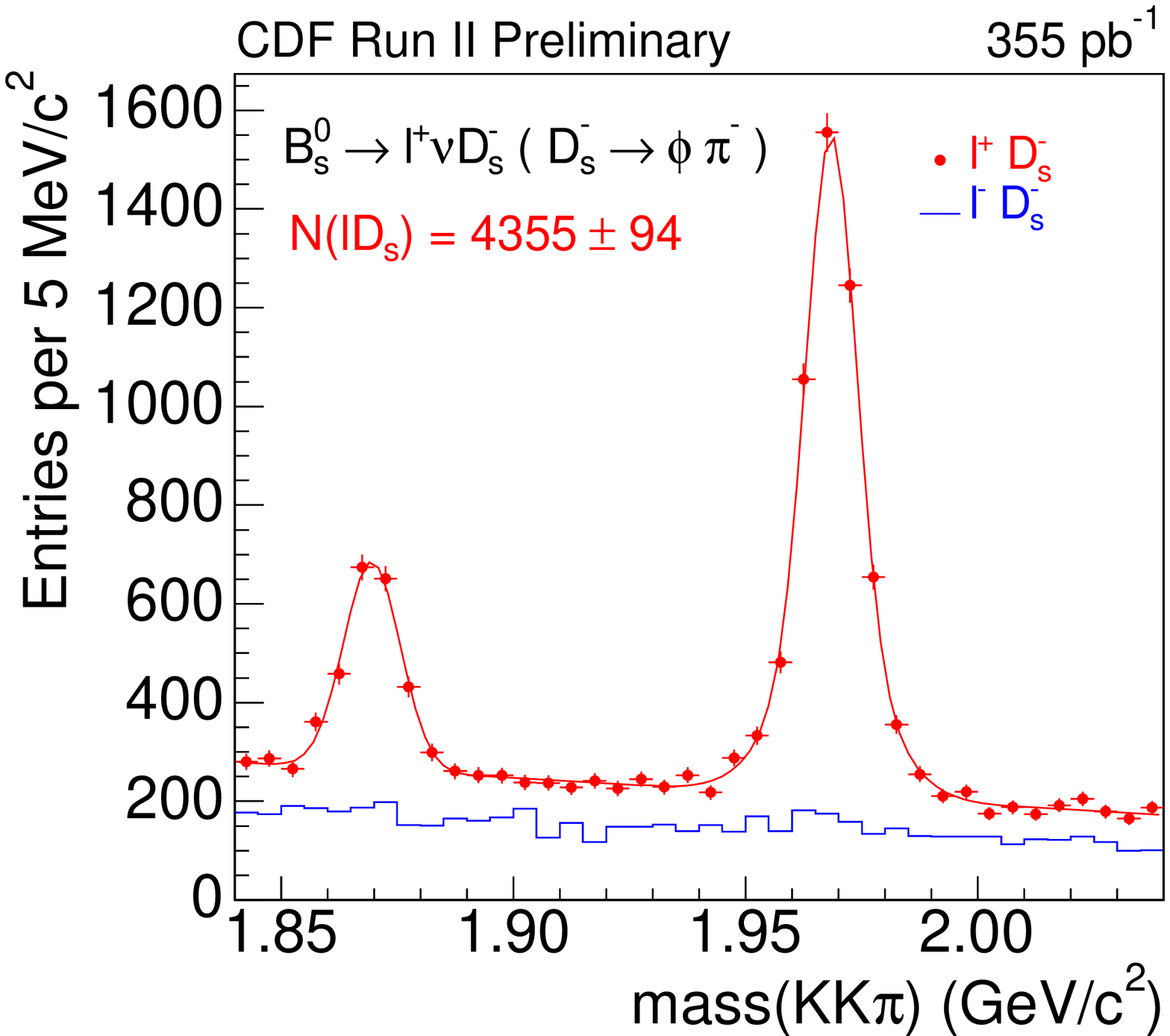}
\vspace{-20pt}
\caption{\BslDs,\DsPhipi signal.}
\label{fig:BsSemil}
\end{figure}

 Currently we reconstruct $\BsDspi$ (hadronic analysis) 
and $\BslDs$ channels (semileptonic analysis). In both cases
the $\DsPhipi$,$\DsKstarK$ and $\DsTrepi$ decay modes
are used, where the light meson resonances are detected
in the $\phi\myto K^+K^-$ and $\kstar\myto K^+\pi^-$ channels only.
These modes combined correspond to 4.9\% of the total $D_s$ decay rate.

 \Bs\ meson signals are identified as sharp peaks 
either in the reconstructed \Bs\ mass for hadronic decays 
or in the $D_s$ mass for opposite sign combinations of a
lepton ($\mu$ or $e$) and a $D_s^\pm$ meson candidate.

 $B$ meson candidates are reconstructed starting from 
charged tracks and leptons matched to trigger objects satisfying
on-line trigger selections, in order to accurately model 
trigger biases to the signal proper time 
distribution through Monte Carlo (MC) simulations. 
To ensure optimal vertexing and momentum resolution 
only tracks reconstructed both in the central drift chamber and in the 
silicon vertex detector with a minimum transverse momentum of 400 \MeVc\ 
are used. A secondary/tertiary vertex fit is performed on all 
possible track combinations using a mass constraint on 
$D_s$ mesons involved in hadronic decays. Further selection criteria 
have been designed to optimize $S/\sqrt(S+B)$, 
using MC simulations for signal and sideband data 
for background. 

 Hadronic decays are selected requiring a significantly 
displaced secondary vertex and a tertiary vertex consistent with a downstream 
decay of a charmed meson. Selecting high probability vertex fit reduces 
combinatorial background 
and a minimum $P_T$ on the pion from \Bs\ decay 
is required
to reduce background from $D_s$ production in $\ppbar \myto c X$ 
events combined with a fragmentation track.
Remaining background sources are combinatorics and 
partially reconstructed $B$ decays. 
The former is modeled by exponential functions and fixed by fits 
to the high mass \Bs\ sideband while the latter 
appear as a well separated bump to the low mass side of the signal peak. 
From MC simulation we derive shapes for the 
contributions from decays like $\Bs\myto D_s^\ast \pi 
\myto D_s  (\gamma,\pi^0)\pi^+$ or $\Bs\myto D_s \rho^+ 
\myto D_s \pi^+ (\pi^0)$ where the $\gamma$ or $\pi^0$ escape 
undetected. Other backgrounds considered 
are Cabibbo-suppressed $\Bs\myto D_s K$
decays (whose contribution is fixed to the 
analogous \Bd\ decay rate) and reflections 
from b-baryon and \Bd\ decays whose shapes and relative
contributions are fixed to those expected from simulation. 

\begin{table}[t]
\caption{\Bs\ yield and S/B in 360 \ipb\ of \cdfii\ data.}
\label{table:sample}
\newcommand{\m}{\hphantom{$-$}}
\newcommand{\cc}[1]{\multicolumn{1}{c}{#1}}
\renewcommand{\arraystretch}{1.2} 
\begin{tabular}{@{}lll}
\hline
Decay Channel            & Yield      & S/B \\
\hline
$\BsDspi$,$(\Phipi)$     & $526\pm33$ & 1.8 \\
$\BsDspi$,$(\KstarK)$   & $254\pm21$ & 1.7  \\
$\BsDspi$,$(\Trepi)$     & $116\pm18$ & 1.0  \\
\hline
$\BslDs$,$(\Phipi)$     & $4355\pm94$ & 3.1  \\
$\BslDs$,$(\KstarK)$   & $1750\pm83$ & 0.4  \\
$\BslDs$,$(\Trepi)$     & $1573\pm88$ & 0.3  \\
\hline
\end{tabular}\\[2pt]
\end{table}

 Similar selection criteria are used to isolate semileptonic decays.
Further requests on the $K^\ast$ and $\phi$ helicity angle and on the 
the lepton + $D_s$ invariant mass ($2.3  < m_{lD_s} <5\ \GeVcc$) 
reduce combinatorial and $c\bar{c}$ background. An important 
background (of order 20\% of the signal sample) arises from 
$B\myto D_s^{(\ast)} D_{(s)}^{(\ast)} X$ with $D_{(s)}^{(\ast)}\myto l\nu X$ 
decays.
The expected contribution is determined from MC using 
world average branching ratios and considered in the mixing 
analysis with proper lifetime and mixing probabilities.
The background from real prompt $D_s^{(\ast)}$ 
paired with a random track faking a lepton (prompt background) 
is estimated from wrong sign, $l^\pm D_s^\pm$, signal.
Its contribution is about 4\% of the right sign sample. 

 An example of the clean signals observed in hadronic and semileptonic
decays is shown in Figs.~\ref{fig:Bshadronic} and \ref{fig:BsSemil}.
Number of events and $S/B$ values extracted from fits to the 
\Bs\ or $D_s$ invariant mass are displayed in Table~\ref{table:sample}.
\section{Proper Time Reconstruction and Resolution}
\begin{table*}[htb]
\caption{Tagging dilution, efficiency and 
calibrated $\varepsilon D^2$ (\%) with statistical and systematic uncertainty.}
\label{tab:ed2}
\newcommand{\m}{\hphantom{$-$}}
\newcommand{\cc}[1]{\multicolumn{1}{c}{#1}}
\renewcommand{\tabcolsep}{2pc} 
\renewcommand{\arraystretch}{1.2} 
\begin{tabular}{@{}lllll}
\hline
Tagger & $\sqrt{<D^2>}$ & $\varepsilon$ 
                        & $\varepsilon (DS_D)^2 {\mathrm(semil.)}$ 
                        & $\varepsilon (DS_D)^2 {\mathrm(had.)}$ \\ 
\hline
SMT    & 35.9 & 5.0  & $0.557\pm0.047\pm0.034$ & $0.46\pm0.11\pm0.03$ \\
SET    & 29.4 & 3.5  & $0.293\pm0.033\pm0.017$ & $0.18\pm0.06\pm0.02$ \\
JVX    & 16.9 & 9.8  & $0.263\pm0.035\pm0.021$ & $0.14\pm0.07\pm0.01$ \\
JJP    & 11.5 & 14.0 & $0.150\pm0.026\pm0.015$ & $0.11\pm0.06\pm0.01$ \\
JPT    &  5.0 & 52.0 & $0.157\pm0.027\pm0.025$ & $0.24\pm0.09\pm0.01$ \\
\hline
Total  & 35.9 & 5.0  &    $1.429\pm0.093$      & $1.12\pm0.18\pm0.04$ \\
\end{tabular}\\[2pt]
\end{table*}
%
%
 The proper decay time of \Bs\ mesons is reconstructed
using the distance in the transverse plane from the beam axis 
to the decay vertex ($L_{xy}$) and the transverse momentum of the candidates:
$c\tau_{B_s} = L_{xy} \cdot m_{B_s}/P_T$.
For semileptonic decays a K factor is needed to compensate 
for missing momentum carried by neutrinos $ c\tau_{B_s}= K \cdot 
c\tau_{B_s}^\ast$ where $c\tau_{B_s}^\ast$ is reconstructed from 
the lepton + $D_s$ transverse momentum. The K factor distribution 
is derived from MC simulation of inclusive semileptonic decays.

 The time dependent mixing probability is convoluted with 
a Gaussian resolution function using the event-by-event expected
error on $c\tau_{B_s}$. The latter includes a scale factor 
derived studying the proper time resolution function on a 
sample of pseudo-$\Bs$ candidates constructed from a prompt $D_s$ meson 
paired with a fragmentation track. This sample exhibits characteristics
very similar to those of the signal and allows the parameterization of the 
scale factor as a function of several kinematic variables. 
The average $c\tau$ uncertainty for the golden 
$\BsDspi$,$\DsPhipi$ mode
is $\approx 100\ {\rm fs}$ corresponding to 60-70 \um\ resolution on the 
\Bs\ decay length.
Note that the exponential term in eq.~\ref{eq:significance} implies
a 60\% loss of statistical power for 100 $fs$ resolution 
at $\Dms = 15$. 
Improving the proper time measurement 
accuracy, e.g. by using an event-by-event primary vertex, 
will certainly help in a future updated analysis.

\section{Lifetime Measurement}

 Determinations of $B$-meson lifetimes in several exclusive 
$B^+$ and \Bd\ decay channels provide an important check on the 
accuracy of proper time measurement. 
This has been successfully performed~\cite{lifetimes} 
using $B\myto J/\psi K^{(\ast)}$ and semileptonic decays collected 
with a version of the trigger with no
impact parameter requirement on tracks. For the present mixing analysis 
an additional crucial point has been the removal of the bias introduced 
by the trigger and offline selection criteria. A detailed Monte Carlo 
simulation of the trigger, including the effects of the variations 
with time of the beam axis position with respect to the detector and 
of trigger algorithms, has been used to derive the efficiency as a function 
of proper time for each of the decay modes analyzed. This method
has been validated measuring $B^+$ and $\Bd$ lifetime in $D\pi$ modes 
collected by the same trigger used for the mixing analysis. The results
($\tau(B^+)({\rm ps})=1.661\pm0.025\pm0.013$,
$\tau(\Bd)({\rm ps})=1.511\pm0.023\pm0.013$,
$\tau(\Bs)({\rm ps})=1.598\pm0.097\pm0.017$)
are consistent with world averages. The systematics are at the level of 
1\% and demonstrate that the the trigger bias is well under control. 
A similar analysis is currently under way 
using the same semileptonic \Bs\ sample
used here. When completed, it will give the best measurement 
of $\Bs$ lifetime in a flavor specific decay. 
For the time being a result consistent with the world average 
with a statistical uncertainty of 2\% has been reported by CDF.

\section{Flavor Tagging}

 Identifying the initial flavor of the $B$ mesons at hadron colliders
has always been a critical task due to the busy
environment of $b$ events in $\ppbar$ collisions.
To determine the \Bs\ flavor algorithms have been developed 
relying either on the other $b$-hadron in the event, 
Opposite Side (OS) taggers, or on the expected charge correlation 
($\Bs - K^+$) between the \Bs\ flavor and the kaons emerging close to the 
\Bs\ flight direction from the fragmentation and hadronization process,
Same Side Kaon (SSK) tagger. Five different OS taggers are used here: 
\begin{itemize}
\item A Soft Muon Tagger (SMT), based on the detection of 
muons, with $P_T>1.5\ \GeVc$ and pseudorapidity $|\eta| < 1.5$ 
from $b\myto \mu^-X $ decays. 
The dilution is parameterized as a function of a muon
likelihood and muon momentum relative to the jet axis, $P^{rel}_T$, to 
account for the higher dilution expected at higher  $P^{rel}_T$ due 
to lower $b \myto c \myto l^+$ sequential background.
\item A similar algorithm, Soft Electron Tagger (SET), 
uses $b$ decays to electrons with $P_T>2.0\ \GeVc$ 
detected in the central electromagnetic calorimeter.
\item Three different Jet Charge Taggers, 
where a weighted sum of track charges in opposite side 
jets is correlated with the initial $b$-quark charge sign. 
The three algorithms differ on the choice of the tagging jet: 
jet containing an identified
secondary vertex (JVX), jet characterized by at least one displaced track
(JJP) or the highest $P_T$ jet in the event (JPT).
\end{itemize}
 All the taggers have been tuned on an inclusive and almost pure sample 
of $\approx 10^6$ semileptonic $b$ decays from the lepton + SVT trigger. 
The dilution of each different algorithm is determined, 
as a function of several variables, from the 
the trigger lepton charge after correction for mixing and 
$b \myto c \myto l^+$ background. Taggers are ranked in
order of their expected dilutions and only the decision from the 
highest dilution tagger available in any single event is used.

 Since the calibration sample for the dilution may be different 
in certain aspects (e.g. $P_T$ spectra) from the signal sample 
a scale factor ($S_D$) is allowed for each of the 5 tagging algorithms. 
The $S_D$ factors for the semileptonic and hadronic analysis are 
fixed by measuring the \Bd\ oscillation frequency and its amplitude
using respectively the high statistics $B_{d,u}\myto l D \nu X$ and 
$B_{d,u}\myto J/\psi K^{(\ast)}/B_{d,u}\myto D^{(\ast)}\pi$ samples. 
A summary of the tagging figure of merit $\varepsilon D^2$ is reported
in Tab.~\ref{tab:ed2}. The scale factors are consistent 
with 1 within the precision allowed by the present statistics.

\section{Results and Perspectives}

 Following~\cite{Moser} an amplitude scan is performed,
where, at each value of the mixing frequency, an amplitude $A$
is fitted from the data. The amplitude $A$ should be consistent 
with 0 for mixing 
frequencies far from the true value of \Dms\ and with 1 in its vicinity. 
This method allows for a straightforward 
way to combine several measurement (semileptonic and hadronic) as well
as results from different experiments. 

We determined systematic 
uncertainties on the amplitude $A$
using large toy MC experiments to minimize the effect 
of the limited statistics. The biggest effects on systematic
uncertainties at high value of \Dms\ come from the level and 
the dilution parameterization of the prompt background 
for the semileptonic analysis (contributing 
$\sigma_A\simeq0.15\ @\ \Dms=20 \psinv$) and the dilution scale factor for 
the hadronic analysis ($\sigma_A\simeq0.3\ @\ \Dms=20 \psinv$).
It should be noticed that the latter will certainly 
be reduced with more data since it is limited by the 
available number of $B_{d,u}$ hadronic decays. In any case 
the combined systematic 
uncertainties are much smaller than the statistical uncertainties
on the present measurements, for any value of the mixing frequency.

\begin{figure}[t]
\includegraphics[width=0.98\linewidth]{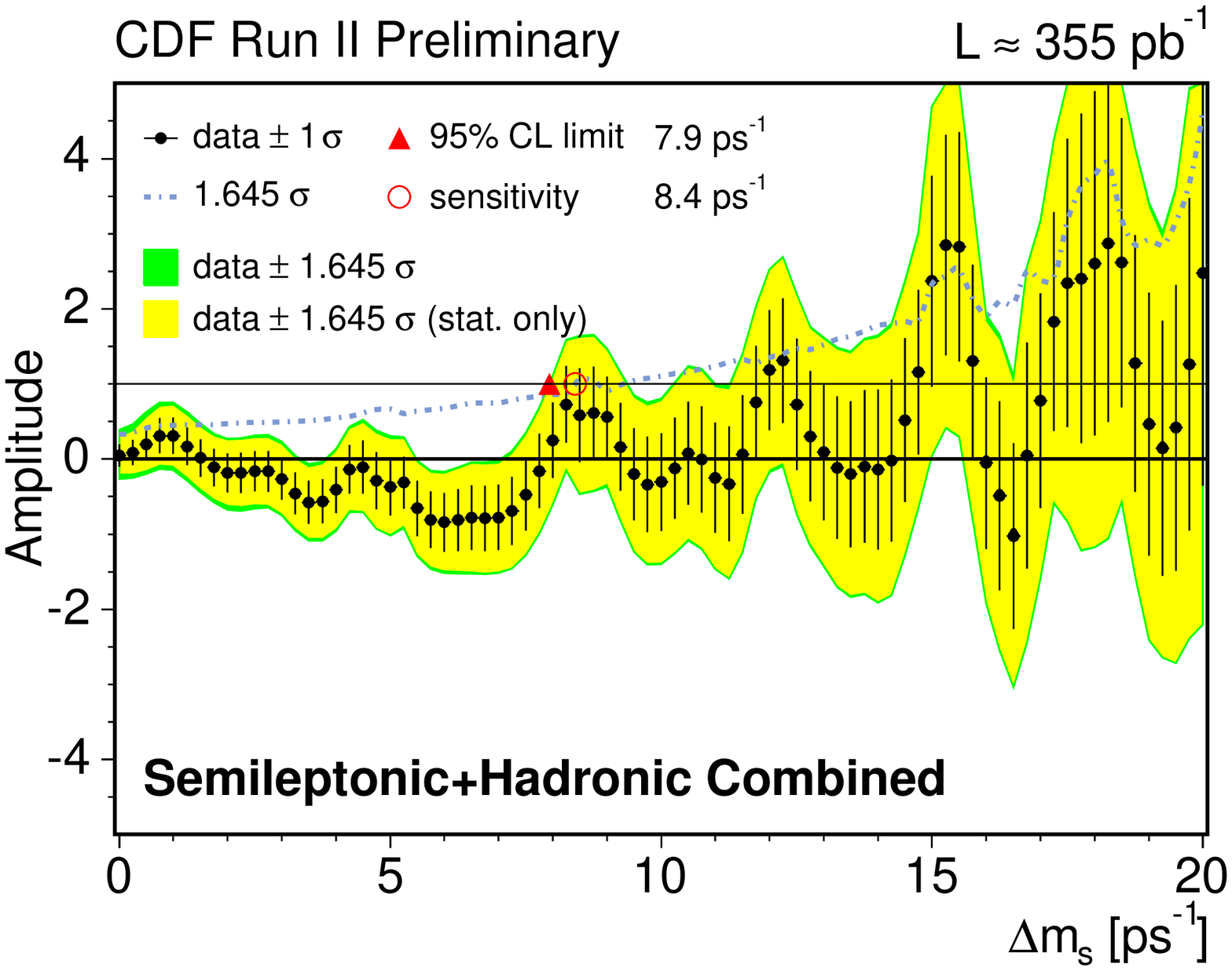}
\vspace{-45pt}
\caption{Amplitude scan for the combined \BslDs\ and \BsDspi\ analysis.}
\label{fig:AmpScans}
\end{figure}


 The present CDF sensitivity is dominated by the semileptonic 
result: $\Dms > 7.7\ \psinv @\ 95\%$ C.L. with a sensitivity of 7.4 \psinv. 
The combined semileptonic and hadronic analysis limit is 
$\Dms > 7.9\ \psinv @\ 95\%$ C.L. while the combined sensitivity 
increases to 8.4\psinv. 

 Combining the CDF mesaurement with previously available 
results does not change the actual limit 
and only marginally improves the sensitivity (from 18.2 to 18.6). 
It can be noticed, however, that the current sensitivity of \cdfii\ data is 
relatively better behaved at higher \Dms\ than most of 
previous experiments due to the valuable hadronic
sample that will dominate the sensitivity in future higher
statistics searches for \Bs\ oscillations.

\end{document}